\documentclass[preprintnumbers,article,amsmath,amssymb,floatfix,10pt,prd,onecolumn,
superscriptaddress,nofootinbib]{revtex4-2}
\usepackage{bm}
\usepackage{amsfonts}
\usepackage{latexsym}
\usepackage[utf8]{inputenc}
\usepackage{graphicx}
\usepackage{mathrsfs}
\usepackage{amsmath}
\usepackage{dirtytalk}
\usepackage{palatino}
\usepackage{mathpazo}
\usepackage{textcomp}
\usepackage{float}
\usepackage{booktabs}
\usepackage{dcolumn}
\usepackage{ragged2e}
\usepackage{hyperref}[green]
\hypersetup{colorlinks,citecolor=blue}
\hypersetup{colorlinks=true,linkcolor=red,filecolor=magenta,    urlcolor=blue}
\usepackage{amsmath}
\usepackage{cleveref}
\usepackage{xcolor}
\usepackage{orcidlink}
\usepackage{epsfig}
\usepackage{caption}
\usepackage{subcaption}
\usepackage[none]{hyphenat}   
\usepackage{commath}
\def\jnl@style{\it}
\def\aaref@jnl#1{{\jnl@style#1}}

\def\aaref@jnl#1{{\jnl@style#1}}

\def\aj{\aaref@jnl{AJ}}                   
\def\apj{\aaref@jnl{ApJ}}                 
\def\apjl{\aaref@jnl{ApJ}}                
\def\apjs{\aaref@jnl{ApJS}}               
\def\apss{\aaref@jnl{Ap\&SS}}             
\def\aap{\aaref@jnl{A\&A}}                
\def\aapr{\aaref@jnl{A\&A~Rev.}}          
\def\aaps{\aaref@jnl{A\&AS}}              
\def\mnras{\aaref@jnl{Mon.~Not.~Roy.~Astron.~Soc.}}             
\def\prd{\aaref@jnl{Phys.~Rev.~D}}        
\def\prc{\aaref@jnl{Phys.~Rev.~C}}  
\def\prl{\aaref@jnl{Phys.~Rev.~Lett.}}    
\def\qjras{\aaref@jnl{QJRAS}}             
\def\skytel{\aaref@jnl{S\&T}}             
\def\ssr{\aaref@jnl{Space~Sci.~Rev.}}     
\def\zap{\aaref@jnl{ZAp}}                 
\def\nat{\aaref@jnl{Nature}}              
\def\aplett{\aaref@jnl{Astrophys.~Lett.}} 
\def\apspr{\aaref@jnl{Astrophys.~Space~Phys.~Res.}} 
\def\physrep{\aaref@jnl{Phys.~Rep.}}      
\def\physscr{\aaref@jnl{Phys.~Scr}}       
\def\commat{\aaref@jnl{Comm.~Math.~Phys.}}              
\def\science{\aaref@jnl{Science}}               
\def\cqg{\aaref@jnl{Classical Quant.~Grav.}}            
\def\jpcs{\aaref@jnl{JPCS}}                                     
\def\ijmpd{\aaref@jnl{Int.~J.~Mod.~Phys.~D}}                    
\def\grg{\aaref@jnl{Gen.~Relat.~Gravit.}}               
\def\rpp{\aaref@jnl{Rep.~Prog.~Phys.}}          
\def\npa{\aaref@jnl{Nucl.~Phys.~A}}        
\def\lrr{\aaref@jnl{Living Rev.~Rel.}}                   
\def\jcap{\aaref@jnl{J.~Cosmology Astropart.~Phys.}}    
\def\rmp{\aaref@jnl{Rev.~Mod.~Phys.}}   
\def\epjc{\aaref@jnl{Eur.~Phys.~J.~C}} 
\def\plb{\aaref@jnl{~Phy.~Lett.~B}} 
\def\mpla{\aaref@jnl{Mod.~Phy.~Lett.~A}} 
\def\arxiv{\aaref@jnl{arxiv.org}}

\allowdisplaybreaks[1]

\begin{document}
\color{black}       

\title{Parameterized Deceleration in $f(Q,C)$ Gravity: A Logarithmic Approach}

\author{S. R. Bhoyar\orcidlink{0000-0001-8427-4540}}
\email{drsrb2014@gmail.com}
\affiliation{Department of Mathematics, Phulsing Naik Mahavidyalaya, Pusad-445216 Dist. Yavatmal, (India).}

\author{Yash B. Ingole\orcidlink{0009-0006-7208-1999}}
\email{ingoleyash01@gmail.com}
\affiliation{Department of Mathematics, Phulsing Naik Mahavidyalaya, Pusad-445216 Dist. Yavatmal, (India).}
%

\begin{abstract}

\textbf{Abstract:} 
This study explores a novel logarithmic parameterization of the deceleration parameter within the \( f(Q, C) \) gravity framework, incorporating a nonlinear functional form \( f(Q, C) = \gamma_1 Q^n + \gamma_2 C \), where \( Q \) and \( C \) denote the nonmetricity scalar and boundary term respectively and $n>1$. This approach provides a distinctive perspective on the accelerated expansion of the universe without resorting to exotic fields. Using Observational Hubble Data  ($OHD$) measurement and the $Pantheon+SH0ES$ Type Ia supernovae dataset, the model parameters were restricted by a $\chi^2$ minimization technique. The analysis reveals a transition from deceleration to acceleration in the expansion history of the universe, with transition redshifts \( z_t \approx 0.98 \) ($OHD$) and \( z_t \approx 0.76 \) ($Pantheon+SH0ES$). The model demonstrates consistency with observations, offering insights into the dynamics of dark energy and alternative gravity theories while effectively modeling cosmic evolution across epochs.
 
\end{abstract}

\maketitle

\date{\today}


\section{Introduction}
\label{section 1}

Recent observations have significantly altered our understanding of the universe, revealing that its expansion is currently accelerating \cite{riess1998observational,perlmutter2003dark}. Evidence from various sources, including Type Ia Supernovae (SNeIa), Cosmic Microwave Background (CMB) radiation, and Baryon Acoustic Oscillations (BAO) \cite{clifton2012modified,riess2019large}, has consistently indicated this acceleration, implying the presence of an energy form with significant negative pressure, commonly known as Dark Energy (DE), which contributes almost 70\% to the current energy budget of the universe. Several DE models have been proposed to explain this phenomenon. The cosmological constant ($\Lambda$), which corresponds to a simple Equation of State (EoS) $\omega=-1$, indicates a constant energy density throughout space and time, which is the simplest and most widely accepted. However, the $\Lambda CDM$ model faces critical challenges, such as the fine-tuning problem \cite{weinberg1989cosmological}, the coincidence problem \cite{steinhardt1999cosmological}, and the age problem \cite{copeland2006dynamics}, which call for alternative explanations for the nature and origin of DE. In response to these issues, scalar field models both canonical and noncanonical have gained prominence because they provide a more dynamic and versatile framework for describing cosmic evolution. Over the past decade, numerous DE models, including quintessence, K-essence, phantom energy, tachyon fields, and Chaplygin gas, have been explored as potential candidates for explaining cosmic acceleration (see \cite{amendola2010dark} for a comprehensive review). Despite this progress, a universally accepted and definitive DE model remains difficult to detect.

These unresolved issues have encouraged the exploration of modified gravity theory, which aims to provide alternative explanations for the accelerated expansion of the universe. Rather than relying on exotic energy components such as DE, these theories propose that cosmic acceleration could arise from modifications to the fundamental laws of gravity. In the following passage, we explore the various modified gravity models, focusing on how they extend the standard framework of General Relativity (GR) and offer new insights into the dynamics of the cosmos.

Modified theories of gravity are advanced frameworks designed to extend or refine the GR to address phenomena that the original theory cannot fully explain. These modifications attempt to provide a deeper insight into the fundamental nature and dynamics of the universe by introducing innovative mathematical concepts and formulations. Some prominent examples of these theories include: \(f(R)\) gravity, which extends the Ricci scalar (\(R\)) \cite{buchdahl1970non,starobinsky1980new}. \(f(T)\) gravity, formulated using the torsion scalar (\(T\)) \cite{ferraro2007modified}. \(f(G)\) gravity, incorporating the Gauss-Bonnet term (\(G\)) \cite{nojiri2005modified}. \(f(R,\mathbb{T})\) gravity, which combines the scalar \(R\) and the trace of the energy-momentum tensor (\(\mathbb{T}\)) \cite{harko2011f}. \(f(R,G)\) gravity, uniting the scalar \(R\) with the term \(G\) \cite{bamba2010finite,de2012stability}. \(f(Q)\) gravity, focused on the nonmetricity scalar (\(Q\)) \cite{jimenez2018coincident}. \(f(Q,\mathbb{T})\) gravity, which merges the scalar \(Q\) with the tensor \(\mathbb{T}\) \cite{xu2019f}. These theories provide a foundation for exploring alternative perspectives on gravitational phenomena and cosmic evolution. Some studies have explored alternative methods to explain cosmic acceleration without assuming specific functional forms for gravity modifications. In particular, a model-independent technique proposed in \cite{capozziello2022model} offers a way to reconstruct cosmological evolution directly from observational data, independent of the underlying gravity theory. This approach provides a powerful tool to analyze cosmic acceleration without relying on specific theoretical frameworks. In contrast, $f(Q, C)$ gravity (which we consider in this research work) explicitly incorporates nonmetricity and the boundary term $C$, offering a more geometrically motivated explanation of cosmic evolution. Although model-independent methods offer flexibility, they do not provide direct insight into the fundamental nature of gravity modifications.

In recent years, alternative theories of gravity have been extensively studied as potential explanations for the accelerated expansion of the universe. Among these, $f(R)$ and $f(T)$ gravity has emerged as a feasible extension of GR, demonstrating the ability to produce late-time cosmic acceleration without introducing scalar fields. $f(R)$ gravity modifies the Einstein-Hilbert action by generalizing the Ricci scalar $R$, allowing dynamical cosmic evolution driven purely by curvature corrections \cite{capozziello2014cosmographic}. Similarly, $f(T)$ gravity reformulates gravity using torsion instead of curvature, where specific functional choices of the Torsion scalar $T$ naturally lead to a transition from deceleration to acceleration \cite{pawar2025perfect}. Both theories provide purely geometric mechanisms for explaining cosmic acceleration, making them compelling alternatives to DE models \cite{mishra2025cosmography}. In this context, $f(Q,C)$ gravity, which is based on the nonmetricity scalar $Q$ and the boundary term $C$, offers another framework in which the expansion history of the universe can be successfully described without requiring exotic scalar fields. The advantage of $f(Q, C)$ gravity lies in its ability to unify different geometric contributions while maintaining consistency with observational data. This work explores the role of $f(Q, C)$ gravity in driving an expansion of the universe and compares its transition behavior with that observed in other modified gravity theories.

A groundbreaking theoretical framework called \(f(Q,C)\) gravity has been introduced, providing new insights into DE and the  accelerating expansion of the universe. This theory investigates the nonlinear interplay between the nonmetricity scalar \(Q\) and the boundary term \(C\), offering a potential explanation for late-time cosmic acceleration without the need for exotic fields or $\Lambda$. The \(f(Q,C)\) gravity model is an extension of the \(f(Q)\) gravity framework, enhancing the nonmetricity-based approach by introducing a second scalar term, \(C\), which depends both on the Hubble parameter and its time derivative. The additional term \(C\), defined as \(C = 6(3H^2 + \dot{H})\), is directly related to the dynamic aspects of cosmic expansion, particularly the rate of change in the Hubble parameter over time \cite{bahamonde2023teleparallel,de2023non,capozziello2023role,gadbail2023cosmological}. This feature sets \(f(Q,C)\) gravity apart, as it combines the geometric property of nonmetricity (represented by \(Q\)) with the evolution of cosmic expansion (captured by \(C\)). By incorporating this unique interplay between \(Q\) and \(C\) within the gravitational action, \(f(Q,C)\), gravity is capable of effectively describing both the decelerating and accelerating phases of the universe. Incorporation of \(C\) in \(f(Q,C)\) gravity introduces new gravitational effects that can be tested with the help of observational data from CMB, Large Scale Structure (LSS) and SNeIa.  In this framework, \(Q\) measures the deviations in the metric during parallel transport, distinguishing it from GR, which assumes a symmetric and torsion-free connection. The boundary term \(C\), which emerges from the interaction between torsion-free and curvature-free connections, ensures the dynamic equivalence of the model to GR  under specific conditions, allowing seamless transitions between different geometric representations of gravity. Additionally, \(C\) introduces extra degrees of freedom that significantly impact the behavior of gravitational fields, particularly on cosmological scales. In this work, the functional form of $f(Q,C)$ gravity is employed as 
\begin{equation*}
f(Q,C)=\gamma_1Q^n+\gamma_2C,   
\end{equation*}
where $\gamma_1$, and $\gamma_2$ are constants; To study the nonlinear nature of $f(Q,C)$ gravity, we adopt the condition that, integer  $n>1$. The reason for considering the nonlinear functional form is explained in \hyperref[section_2]{Section II}. \cite{de2023non,SAMADDAR2024116643,capozziello2023role,maurya2024quintessence,maurya2024modified,chandra2024transit,maurya2024cosmology}. These studies explain the recent work on $f(Q,C)$.

Cosmological observations suggest that the observed accelerated expansion of the universe is a relatively recent phenomenon. During early epochs, particularly in the matter-dominated era, when DE was absent or had a negligible effect, the universe must have experienced a decelerated phase to enable the formation of cosmic structures, as gravitational forces held matter together. Consequently, a comprehensive cosmological model must encompass decelerated and accelerated phases of expansion to accurately represent the evolutionary history of the universe. In this context, the deceleration parameter serves as a crucial tool. One of the most common ways to achieve such a scenario is through parameterization of the deceleration parameter, expressed as a function of the scale factor (\(a\)), redshift (\(z\)), or cosmic time (\(t\)) (refer to  \cite{pr1,pr2,pr3,pr4,pr5,pr6}). On the other hand, nonparametric methods, which directly derive the evolution of the universe from observational data without assuming specific parameterizations, offer advantages by avoiding constraints on cosmological quantities; they also have certain limitations \cite{pr7,pr8,pr9,pr10,pr11}. To date, no theoretical model has been found that can fully describe the entire evolution of the universe. Therefore, adopting a parametric approach remains a practical choice.

There are several key reasons for adopting such a parameterization. First, it provides a flexible yet controlled approach to studying cosmic evolution, allowing for the exploration of complex dynamics without the need to solve intricate differential equations. Second, parameterized forms are well suited for observational studies, as they enable a direct comparison between theoretical predictions and data from SNeIa, CMB, and BAO \cite{riess1998observational,boehm2003t}.  Third, they allow for a more straightforward understanding of how the deceleration parameter evolves over cosmic time, capturing key features of cosmological models in terms of a few well-defined parameters \cite{padmanabhan2003cosmological}. This structured approach enables a detailed investigation of the DE role in cosmic dynamics and its potential implications for modified gravity theories, providing valuable insights into the expansion history of the universe.

In the \( f(Q, C) \) gravity framework, parameterizing the deceleration parameter $q(z)$ is particularly beneficial for investigating the implications of cosmic acceleration theories. Unlike the static cosmological constant model, which requires precise fine-tuning. \( f(Q, C) \) gravity integrates nonmetricity and dynamic terms that evolve with cosmic time. This combination enables the model to describe the transition between the decelerating and accelerating phases of the expansion of the universe. Parameterization of $q(z)$ within this framework provides a structured approach to compare theoretical predictions with empirical data from multiple cosmological probes, such as SNeIa, CMB, and BAO. In particular, many of these parameterizations diverge in the far future, whereas others are only valid for low redshift values (\( z \ll 1 \)) \cite{gong2007reconstruction,cunha2008transition,cunha2009kinematic,santos2011current,nair2012cosmokinetics}.

Motivated by these considerations, our study adopts the parameterization of a specific form of the deceleration parameter: 
\begin{equation*}
   q(z) = q_0 + q_1 \left[ \frac{\ln[\alpha + z]}{1 + z} - \beta \right], 
\end{equation*}
where $q_0, q_1,\alpha,$ and $\beta$ are arbitrary model parameters. The logarithmic term is pivotal in capturing the dynamics of \( q(z) \). The inclusion of this term ensures a smooth and controlled evolution of the deceleration parameter, facilitating a gradual transition between the deceleration and acceleration phases rather than an abrupt shift. This smooth evolution is essential for modeling subtle dynamic changes in the expansion history of the universe \cite{aviles2013cosmographic}. Moreover, at high redshifts (\( z \gg 1 \)), the logarithmic term evolves more gradually than the linear terms, avoiding unphysical divergences and maintaining consistency with early-universe observations \cite{copeland2006dynamics}. This parameterization significantly enhances the study of the  expansion of the universe by providing a unified framework to model its evolutionary dynamics. It captures the intricate transition between matter-dominated cosmic deceleration and DE-driven acceleration within a single mathematical expression. Adjusting the model parameters allows for a detailed investigation of the interplay between various cosmic components in different epochs, offering deeper insights into the nature of DE and the mechanisms underlying the accelerated expansion of the universe. Furthermore, the controlled and smooth behavior of the logarithmic term ensures alignment with observational constraints from both late-time and early-time cosmological data, making it a robust and versatile tool for understanding the dynamics of cosmic evolution.

This article examines the parameterization of the deceleration parameter $q(z) = q_0 + q_1 \left[ \frac{\ln[\alpha + z]}{1 + z} - \beta \right]$ as a function of  the redshift  \( z \). The chosen parameterization exhibits the desired characteristic of transitioning from a decelerating to an accelerating phase. The study investigates the Friedmann–Lemaître–Robertson–Walker (FLRW) universe within the framework of \( f(Q,C) \) gravity, adopting the functional form $f(Q,C)=\gamma_1Q^n+\gamma_2C$. To determine the best-fit values of the model parameters, the chi-square (\( \chi^2 \)) minimization technique is employed. By comparing theoretical predictions with observational data, this study identifies the parameter set that was most closely aligned with empirical evidence, facilitated by statistical analysis. The paper is organized as follows. In \hyperref[section_2]{Section II}, we outline the fundamental formalism of \( f(Q,C) \) gravity and derive the field equations for the FLRW metric. \hyperref[section_3]{Section III} introduces the parametric form of the deceleration parameter and determines the corresponding Hubble solution. In \hyperref[section_4]{Section IV}, we apply Bayesian analysis to observational datasets, including data from $OHD$ and $Pantheon+SH0ES$, to constrain the free parameters of the model. \hyperref[section_5]{Section V} examined the evolutionary trajectories of energy density, pressure, the EoS parameter, the statefinder parameters, and the $Om$ diagnostic to demonstrate the accelerating behavior of the universe. In \hyperref[section_6]{Section VI}, we compare our approach in $f(Q,C)$ gravity with model-independent reconstruction techniques, highlighting the advantages of our parameterization in capturing late-time cosmic acceleration while maintaining a theoretical foundation. Finally, \hyperref[section_7]{Section VII} summarizes and concludes the results.

\section{$f(Q,C)$ gravity and field equations}\label{section_2}

In GR, the Levi-Civita connection $\mathring\Gamma^\lambda_{\mu\nu}$ satisfies two important properties: metric compatibility and torsion-free behavior. However, in symmetric teleparallel geometry, these restrictions are removed. Instead, the theory employs an affine connection \(\Gamma^\lambda_{\mu\nu}\) free of torsion and curvature , which is symmetric in its lower indices, justifies the term symmetric teleparallelism. The nonmetricity tensor $Q_{\lambda\mu\nu}$ signifies that the affine connection is not consistent with the metric, which is described by

\begin{equation}\label{1}
    Q_{\lambda\mu\nu} = \nabla_\lambda g_{\mu\nu} = \partial_\lambda g_{\mu\nu} - \Gamma^\beta_{\lambda\mu} g_{\beta\nu} - \Gamma^\beta_{\lambda\nu} g_{\beta\mu} \neq 0.
\end{equation}

The affine connection can be expressed as a combination of the Levi-Civita connection $\mathring\Gamma^\lambda_{\mu\nu}$ and an additional term, the disformation tensor, \( L^\lambda_{\mu\nu} \)  as, 
\begin{equation}\label{2}
    \Gamma^\lambda{}_{\mu\nu} = \mathring\Gamma^\lambda{}_{\mu\nu} + L^\lambda{}_{\mu\nu},
\end{equation}

where
\begin{equation}\label{3}
    L^\lambda{}_{\mu\nu} = \frac{1}{2}(Q^\lambda{}_{\mu\nu} - Q_\mu{}^\lambda{}_\nu - Q_\nu{}^\lambda{}_\mu).
\end{equation}

Two important nonmetricity vectors are derived from the nonmetricity tensor:
\begin{equation}\label{4}
    Q_\mu = g^{\nu \lambda}Q_{\mu\nu\lambda}=Q_\mu{}^\nu{}_\nu, \quad \quad\tilde{Q}_\mu = g^{\nu \lambda}Q_{\nu\mu\lambda}=Q_{\nu\mu}{}^\nu.
\end{equation}

Similar vectors are also defined from the disformation tensor:
\begin{equation}\label{5}
    L_\mu = L_\mu{}^\nu{}_\nu, \quad  \quad \quad \tilde{L}_\mu = L_{\nu\mu}{}^\nu.
\end{equation}

To connect nonmetricity with gravitational dynamics, a superpotential tensor \( P^\lambda{}_{\mu\nu} \), also called the conjugate of nonmetricity, is introduced. It is expressed as:
\begin{equation}\label{6}
   P^\lambda{}_{\mu\nu} = \frac{1}{4} \left(-2L^\lambda{}_{\mu\nu} + Q^\lambda g_{\mu\nu} - \tilde{Q}^\lambda g_{\mu\nu} - \delta^\lambda_\mu Q_{\nu} \right), 
\end{equation}

where parentheses denote symmetrization over indices. Using \( P^\lambda{}_{\mu\nu} \), the nonmetricity scalar \( Q \) is defined as:
\begin{equation}\label{7}
    Q = Q_{\alpha\beta\gamma} P^{\alpha\beta\gamma}.
\end{equation}

Due to the torsion-free and curvature-free properties, certain geometric relationships hold:
\begin{equation}\label{8}
    \mathring{R}_{\mu\nu} + \mathring{\nabla}_\alpha L^\alpha{}_{\mu\nu} - \mathring{\nabla}_\nu \tilde{L}_\mu + \tilde{L}_\alpha L^\alpha{}_{\mu\nu} - L_{\alpha\beta\nu} L^{\beta\alpha}{}_{\mu} = 0,
\end{equation}

\begin{equation}\label{9}
    \mathring{R} + \mathring{\nabla}_\alpha (L^\alpha - \tilde{L}^\alpha) - Q = 0.
\end{equation}

The boundary term \( C \) is then introduced to relate the nonmetricity scalar \( Q \) to the Ricci scalar \\$\mathring{R}$ of Levi-Civita geometry. From the above relation, we can also define the boundary term as,
\begin{equation}\label{10}
    C = \mathring{R} - Q = -\mathring{\nabla}_\alpha  (Q^\alpha - \tilde{Q}^\alpha),
\end{equation}
where the expression $Q^\alpha - \tilde{Q}^\alpha=L^\alpha - \tilde{L}^\alpha$ indicates that the boundary term \( C \) represents the difference between the nonmetricity vectors. This formulation highlights that \( C \) not only encapsulates the difference between \( \mathring{R} \) and \( Q \) but also illustrates their relationship through the divergence of the nonmetricity vector differences. Fundamentally, this establishes a link between the geometric structure of spacetime and the behavior of the nonmetricity vectors, demonstrating how variations in nonmetricity influence the curvature within the framework of the theory.

The action of \( f(Q, C) \) gravity is defined as:
\begin{equation}\label{11}
    S = \int \left[\frac{1}{2\kappa} f(Q,C) + \mathscr{L}_m  \right] \sqrt{-g} \, d^4x,
\end{equation}

where \( f(Q,C) \) is a general function of the nonmetricity scalar \( Q \) and the boundary term \( C \), \(\kappa=\frac{8\pi G}{c^4}\) is the coupling constant, and \( \mathscr{L}_m \) represents Lagrangian matter and $\sqrt{-g}=det(e^A_\mu)=e$.

Varying the action to the metric yields the field equations:
\begin{equation}\label{12}
\begin{split}
    \kappa T_{\mu\nu} = -\frac{f}{2} g_{\mu\nu} + \frac{2}{\sqrt{-g}} \partial_\lambda \left(\sqrt{-g} f_Q P^\lambda{}_{\mu\nu} \right) + \left(P_{\mu\alpha\beta} Q_\nu{}^{\alpha\beta} - 2 P_{\alpha\beta\nu} Q^\alpha{}^\beta{}_\mu \right) f_Q  + \\
\left(\frac{C}{2} g_{\mu\nu} - \mathring{\nabla}_\mu  \mathring{\nabla}_\nu  + g_{\mu\nu} \mathring{\nabla}^\alpha  \mathring{\nabla}_\alpha  - 2 P^\lambda{}_{\mu\nu} \partial_\lambda \right) f_C.
\end{split}
\end{equation}
For a detailed demonstration of this equation, we refer to \cite{samaddar2024constraining}. The covariant form is given as
\begin{equation}\label{13}
\begin{split}
    \kappa T_{\mu\nu} = -\frac{f}{2}+2P^\lambda{}_{\mu\nu}\nabla_\lambda(f_Q-f_c) + \Big(\mathring{G}_{\mu\nu}+\frac{Q}{2}g_{\mu\nu}\Big)f_Q +\\
    \Big(\frac{C}{2}g_{\mu\nu}-\mathring{\nabla}_{\mu}\mathring{\nabla}_\nu+g_{\mu\nu}\mathring{\nabla}^{\alpha}\mathring{\nabla}_\alpha\Big)f_C,
    \end{split}
\end{equation}

where $\mathring{G}_{\mu\nu}$ is an Einstein tensor that corresponds to the Levi-Civita connection. An effective energy-momentum tensor \( T_{\mu\nu}^\text{eff} \) is introduced to simplify the equations. This tensor accounts for the geometric modifications and generates additional terms, replicating the effects of DE:

\begin{equation}\label{14}
\begin{split}
    T_{\mu\nu}^\text{eff} = T_{\mu\nu} + \frac{1}{\kappa} \left[\frac{f}{2} g_{\mu\nu} - 2P^\lambda{}_{\mu\nu} \nabla_\lambda (f_Q - f_C) -  
\frac{Q f_Q}{2} g_{\mu\nu} - 
\left( \frac{C}{2} g_{\mu\nu} -
\mathring{\nabla}_\mu \mathring{\nabla}_\nu + g_{\mu\nu} \mathring{\nabla}^\alpha \mathring{\nabla}_\alpha \right) f_C \right].
\end{split}
\end{equation}

Using the above equation, we derive an equation analogous to that in GR  as follows:
\begin{equation}\label{15}
    \mathring{G}_{\mu\nu}=\frac{\kappa}{f_Q}T_{\mu\nu}^\text{eff}.
\end{equation}

We consider the energy-momentum tensor of a perfect fluid to be:

\begin{equation}\label{16}
    T_{\mu\nu}=pg_{\mu\nu}+(\rho+p)u_\mu u_\nu,
\end{equation}
where $\rho$ is the energy density, $p$ is the pressure and $u^\mu$ represents the four velocities of the fluid.

In this theory, where the affine connection is considered an independent entity, the connection field equation is derived by varying the action with respect to the affine connection.

\begin{equation}\label{17}
    (\nabla_\mu-\tilde{L}_\mu) (\nabla_\nu-\tilde{L}_\nu)[4(f_Q-f_C)P^{\mu\nu}{}_\lambda+\nabla_\lambda{}^{\mu\nu}]=0,
\end{equation}
where 
\begin{equation*}
    \nabla_\lambda{}^{\mu\nu}=-\frac{2}{\sqrt{g}}\frac{\delta(\sqrt{-g}\mathscr{L}_m )}{\delta  \Gamma^\lambda{}_{\mu\nu}}
\end{equation*}

The cosmological principle states that on large scales the universe is homogeneous and isotropic. This assumption leads to the widely used FLRW metric, which describes a spatially flat universe. In Cartesian coordinates, the FLRW metric is expressed as:
\begin{equation}\label{18}
    ds^2 = -dt^2 + a^2(t)(dx^2 + dy^2 + dz^2),
\end{equation}

where \( a(t) \) is the scale factor that determines the expansion of the  universe. The Hubble parameter, which measures the rate of expansion, is defined as $H(t) = \frac{\dot{a}}{a},$ where the overdot (\( \dot{a} \)) represents the derivative for cosmic time \( t \).

As demonstrated in the previous section, the framework of \( f(Q, C) \) gravity introduces an additional effective sector of geometric origin, as expressed in Eqn. \eqref{14}. In a cosmological context, this extra term can be interpreted as an effective DE component characterized by its corresponding energy-momentum tensor.

\begin{equation}\label{19}
    T_{\mu\nu}^\text{DE}=\frac{1}{f_Q}\Big[\frac{f}{2}g_{\mu\nu}-2P^\lambda{}_{\mu\nu}\nabla_\lambda(f_Q-f_C)-\frac{Qf_Q}{2}g_{\mu\nu}-\Big(\frac{C}{2}g_{\mu\nu}-\mathring{\nabla_\mu}\mathring{\nabla}_\nu+g_{\mu\nu}\mathring{\nabla^\alpha}\mathring{\nabla_\alpha}\Big)f_C\Big]
\end{equation}
Fundamental to all modified gravity theories, this additional \( T^\text{DE}_{\mu\nu} \) component essentially produces negative pressure, which drives the late-time acceleration of the universe.

The Lie derivatives of the connection coefficients concerning the generating vector fields of spatial rotations and translation vanish in a symmetric teleparallel affine connection, which is a torsion-free, curvature-free affine connection with both spherical and translational symmetries. In the context of \( f(Q,C) \) gravity, we work with a vanishing affine connection \( \Gamma^\alpha{}_{\mu\nu} = 0 \). To explore this setup in greater detail, refer to \cite{de2024non}, and the following quantities are derived:

\begin{equation}\label{20}
  \mathring{G}_{\mu\nu}=-(3H^2 + 2\dot{H})h_{\mu\nu}+3H^2u_\mu u_\nu,\quad \mathring{R}  = 6(2H^2 + \dot{H}), \quad Q = -6H^2, \quad C = 6(3H^2 + \dot{H}),
\end{equation}
where $u_\nu=(dt)_\nu$, $h_{\mu\nu}=g_{\mu\nu}+u_\mu u_\nu$. The modified Friedmann-like equations in \( f(Q,C) \) gravity are obtained by introducing these quantities in the general field Eq. \eqref{12} as

\begin{equation}\label{21}
    3H^2=\kappa(\rho_m+\rho_{DE}),
\end{equation}
\begin{equation}\label{22}
    -(2\dot{H}+3H^2)=\kappa(p_m+p_{DE}),
\end{equation}

where $\rho_m$ and $p_m$ are the energy density and pressure of the matter sector, respectively. The effective DE density and pressure are defined as

\begin{equation}\label{23}
    \kappa \rho_{DE} =- \frac{f}{2} + 3H^2(1- 2f_Q) + (9H^2 + 3\dot{H}) f_C - 3H \dot{f}_C,
\end{equation}
\begin{equation}\label{24}
    \kappa p_{DE} = \frac{f}{2}-2\dot{H}(1-f_Q)-3H^2(1-2f_Q)+2H\dot{f_Q}- (9H^2 + 3\dot{H}) f_C + \ddot{f}_C,
\end{equation}

where \( f_Q = \partial f / \partial Q \) and \( f_C = \partial f / \partial C \) are partial derivatives of the function \( f(Q,C) \) with respect to \( Q \) and \( C \), respectively. The derivatives \( \dot{f}_C \) and \( \ddot{f}_C \) represent the time derivatives of \( f_C \). These equations generalize the standard FLRW equations of GR by incorporating additional contributions from the dependent terms \( Q \) and \( C \) in \( f(Q,C) \) gravity. The first equation corresponds to the energy constraint, whereas the second governs the dynamics of the universe’s expansion.

\section{Parameterizing the Deceleration Parameter\label{section_3}
}
In general, the  deceleration parameter $q(z)$ in terms of  $H(z)$ is given as 
\begin{equation}\label{25}
    q(z)=-1+\frac{(1+z)}{H(z)}H',
\end{equation}
 where $H'=\frac{dH(z)}{dz}$.
The parameterization of the deceleration parameter significantly influences the nature of the universe's expansion. Some studies used a variety of parametric forms of deceleration parameters in this regard, whereas others looked at nonparametric forms. These techniques have been extensively addressed in the literature to characterize issues with cosmological inquiries, including the Hubble tension, the initial singularity problem, the horizon problem, the all-time decelerating expansion problem, and others \cite{banerjee2005acceleration,cunha2008transition,escamilla2022dynamical}.  Inspired by this finding, we analyze the parametric form of the deceleration parameter in terms of redshift $z$  in this article as:
\begin{equation}\label{26}
    q(z)=q_0+q_1 F(z),
\end{equation}
where $q_0$ and  $q_1$ are free parameters, whereas $F(z)$ is a function of $z$. Several functional forms of  $F(z)$ have been presented \cite{cunha2008transition,turner2002type,riess2004type,gong2006observational,xu2008constraints,del2012three}, which can satisfactorily address several cosmological issues.  However, as already established, some of these parameters lose their power to forecast how the universe will evolve in the future, whereas others are only applicable for $z<<1$.  Moreover, \textit{A.A. Mamon et al.} investigated the divergence-free parameterization of $q(z)$ to study the history of the expansion of the universe \cite{al2016divergence}.  They demonstrated that such a model is more in line with the existing observational constraints for certain model parameter restrictions. Therefore, efforts are being made to find a suitable functional form of $q(z)$ that will work well to address cosmological issues. Inspired by these facts, we adopt a parameterization of the deceleration parameter in this article, which is provided by 

\begin{equation}\label{27}
    q(z)=q_0  +q_1 \Big[\frac{\log[\alpha+z]}{1+z}-\beta\Big],
\end{equation}
where $\alpha$ and $\beta$ are arbitrary model parameters. Eq. \eqref{27} shows two limiting conditions.
\begin{itemize}
    \item At the early epoch, i.e., $z\to \infty$ $\implies$ $q(z)=q_0-\beta q_1$
    \item At the current epoch. i.e., $z=0$ $\implies$ $q(z)=q_0+q_1 \log[\alpha-\beta]$
\end{itemize}

We derive the differential equation by solving Eqns. \eqref{25} and \eqref{27}. In further calculations, we obtain the following results.

\begin{equation}\label{28}
    H(z)=\Bigg[(\alpha+z)^\frac{-q_1}{1+z}\Big(\frac{\alpha+z}{1+z}\Big)^\frac{q_1}{\alpha-1}(1+z)^{1+q_0-\beta q_1}c\Bigg],
\end{equation}

where $c$ is the integrating constant. Furthermore, to eliminate $c$, we assume a boundary condition, i.e., $z=0$. When solving Eq. \eqref{28}, we, obtain
\begin{equation}\label{29}
    H_0= \alpha^\frac{-q_1\alpha}{\alpha-1}c,
\end{equation}
where $H_0$ is the Hubble value/constant at $z=0$. We substituted this value into the  Eq. \eqref{28}, the expression for the Hubble parameter $H(z)$ is obtained as
\begin{equation}\label{30}
    H(z)=H_0\Bigg[\alpha^\frac{q_1\alpha}{\alpha-1}(\alpha+z)^\frac{-q_1}{1+z}\Big(\frac{\alpha+z}{1+z}\Big)^\frac{q_1}{\alpha-1}(1+z)^{1+q_0-\beta q_1}\Bigg].
\end{equation}

Now  we consider the nonlinear $f(Q,C)$ model as:
\begin{equation}\label{31}
    f(Q,C)=\gamma_1Q^n+\gamma_2C,
\end{equation}

where $\gamma_1, \gamma_2$ and $n$ are constants. The choice of a nonlinear form over a linear form of $f(Q,C)$ gravity has been thoroughly justified in the literature \cite{SAMADDAR2024116643}. The dynamical system in $f(Q)$ gravity theory was thought to be analyzed by  \textit{Rana et al.} \cite{rana2024phase} in the form $f(Q)=-Q+\gamma_1Q^n$. In contrast, \textit{D.C. Maurya} examined the $f(Q,C)=\gamma Q^2+\gamma_2 C^2$ form to examine the behavior of quintessence in $f(Q,C)$ gravity theory \cite{maurya2024quintessence}. Several authors have used a variety of nonlinear forms in various gravity theories. Motivated by the models mentioned above, we consider this specific nonlinear form in our computations.

 \section{Data Interpretation}
\label{section_4}
This section outlines the methodologies and selection of observational datasets utilized to constrain the parameters \( H_0 \), \( q_0 \), \( q_1 \), \( \alpha \), and \( \beta \) in the proposed cosmological model. The posterior distributions of these parameters are derived through statistical analysis, specifically employing the Markov Chain Monte Carlo (MCMC) technique. For data analysis, the Python module \textit{emcee} is used.

The probability function \( \mathcal{L} \propto \exp(-\chi^{2}/2) \) is used to optimize the parameter fit, where \( \chi^{2} \) denotes the pseudo-chi-squared function \cite{hobson2010bayesian}. Details of the \( \chi^{2} \) function for various data samples are discussed in the following subsections. The MCMC plot features \( 1\text{-D} \) curves for each model parameter, obtained by marginalizing the remaining parameters. The thick-line curve represents the best-fit value. The diagonal panels of the plot show these \( 1\text{-D} \) distributions, whereas the off-diagonal panels illustrate \( 2\text{-D} \) projections of the posterior probability distributions for parameter pairs. These panels include contours highlighting the regions corresponding to the \( 1\text{-}\sigma \) and \( 2\text{-}\sigma \) confidence levels.

\subsection{Observed Hubble Data}
Accurately determining the expansion rate as a function of cosmic time $t$ is challenging. The Cosmic Chronometer (CC) method offers a different and potentially valuable approach because the expansion rate can be expressed as \( H(z) = \dot{a}/a = -[1/(1 + z)] dz/dt \). In this method, only the Differential Age (DA) progression of the universe, \( \Delta t \), within a specific redshift interval \( \Delta z \), needs to be measured, as \( \Delta z \) is obtained from high-precision spectroscopic surveys. From the ratio \( \Delta z / \Delta t \), an approximate value for \( dz/dt \) can be determined.

\begin{figure}[ht!]
    \centering
    \includegraphics[width=0.5\textwidth]{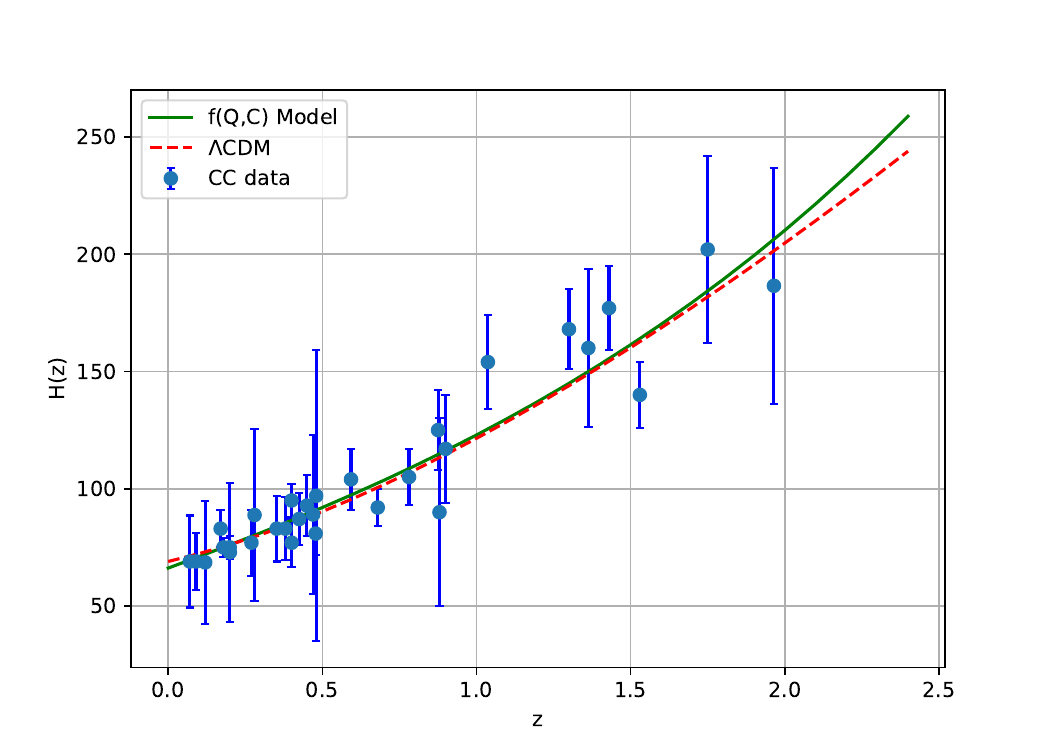} 
    \caption{Error bar plots for 31 data points from the Hubble datasets, together with best-fit plots.}
    \label{I} 
\end{figure}

\begin{figure}[ht]
    \centering
    \includegraphics[width=0.9\textwidth]{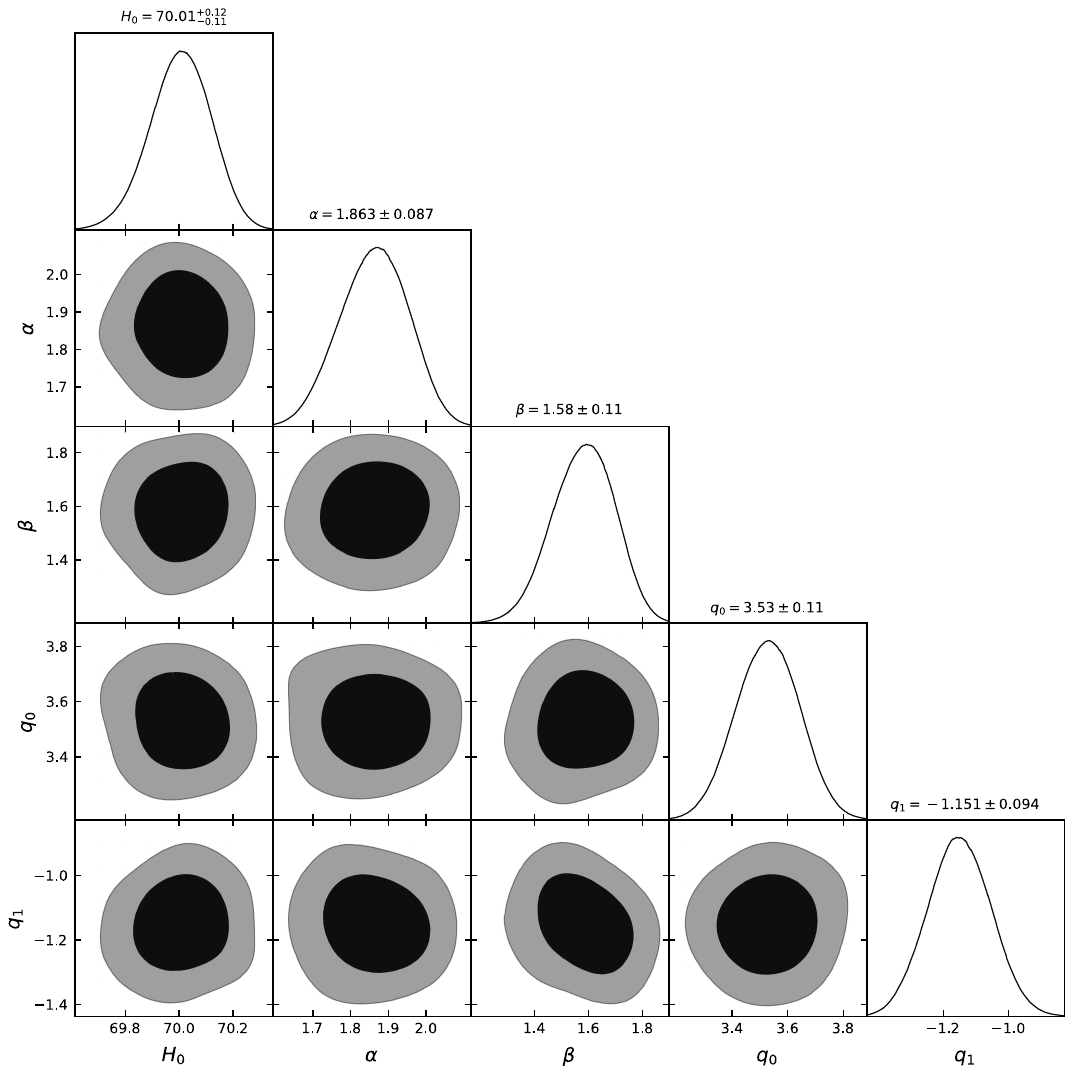} 
    \caption{Marginalized constraints on the coefficients in the expression of  $H(z)$ in Eqn. \ref{30} are shown by using the Hubble sample.}
    \label{II} 
\end{figure}

To estimate the model parameters, we used 31 data points from the \( H(z) \) datasets derived using the DA technique, spanning the redshift range \( 0.07 < z < 2.42 \). The complete list of this dataset is compiled in \cite{singirikonda2020model}. The chi-square function used to deduce the model parameters is as follows:

\begin{equation}\label{32}
    \chi^2_{CC}=\sum_{i=1}^
    {31} \bigg[\frac{H_{i}^{th}(\Theta_{s},z_{i})-H_{i}^{obs}(z_i)}{\sigma_{H(z_i)}}\biggr]^2,
\end{equation}

where \( H^{th} \) and \( H^{obs} \) represent the theoretical and observed values of the Hubble parameter, respectively. The parameter set \( \Theta_s = (H_0, q_0, q_1, \alpha, \beta) \) defines the space of the cosmological background parameter. The standard deviation of the \( i^{\text{th}} \) data point is denoted by \( \sigma_{H(z_i)} \). Figure \ref{I} shows the Hubble parameter profile derived from the CC dataset alongside the behavior predicted by the \( \Lambda \)CDM model. For the MCMC analysis, we employed 100 walkers and 10,000 steps to obtain the fitting results. Contour plots showing the \( 1\text{-}\sigma \) and \( 2\text{-}\sigma \) confidence levels are provided in Figure \ref{II}. Although the model aligns closely with the \( \Lambda \)CDM paradigm at low redshift, noticeable deviations appear at higher redshifts. The marginal values of all the model parameters derived from the Hubble dataset are summarized in Table \ref{tab:I}.

\subsection{\textit{Pantheon+SH0ES} Data}

The discovery of the accelerated expansion of the universe has been significantly advanced through observations of SNeIa. SNeIa has proven to be one of the most powerful tools for investigating the properties of the components driving the universe's rapid evolution. In recent years, numerous compilations of SNeIa data, such as the Joint Light-Curve Analysis (JLA), Pantheon, Pantheon+, Union, Union 2, and Union 2.1 \cite{kowalski2008improved,amanullah2010spectra,suzuki2012hubble,betoule2014improved,scolnic2018complete}, have been published. The $Pantheon+SH0ES$ dataset consists of 1701 light curves from 1550 distinct SNeIa, covering redshifts from \( z = 0.00122 \) to \( 2.2613 \). To constrain the model parameters, the observed and theoretical distance modulus values must be compared. The theoretical distance modulus, \( \mu^{\text{th}}_i \), is expressed as:
\begin{equation}\label{33}
    \mu^{th}_{i}(z,\theta)=5\log D_l(z,\theta)+25,
\end{equation}
where $D_l$ is the dimensionless luminosity distance defined as,
\begin{equation}\label{34}
D_l(z,\theta)=(1+z)\int^z_0 \frac{d\Bar{z}}{H(\Bar{z})}.
\end{equation}

Now, the chi-square function is defined as:
\begin{equation}\label{35}
    \chi^2_{SN}(z,\theta)=\sum_{i,j=1}^{1701} \nabla\mu_{i}(C^{-1}_{SN})_{ij} \nabla\mu_{j},
\end{equation}

\( \nabla\mu_i = \mu^{\text{th}}_i(z, \theta) - \mu^{\text{obs}}_i \) represents the difference between the theoretical and observed distance moduli. The observed distance modulus is denoted as \( \mu^{\text{obs}}_i \), where \( \theta \) defines the parameter space, and \( C_{SN} \) is the covariance matrix \cite{scolnic2022pantheon+}. 

\begin{figure}[ht!]
    \centering
    \includegraphics[width=0.5\textwidth]{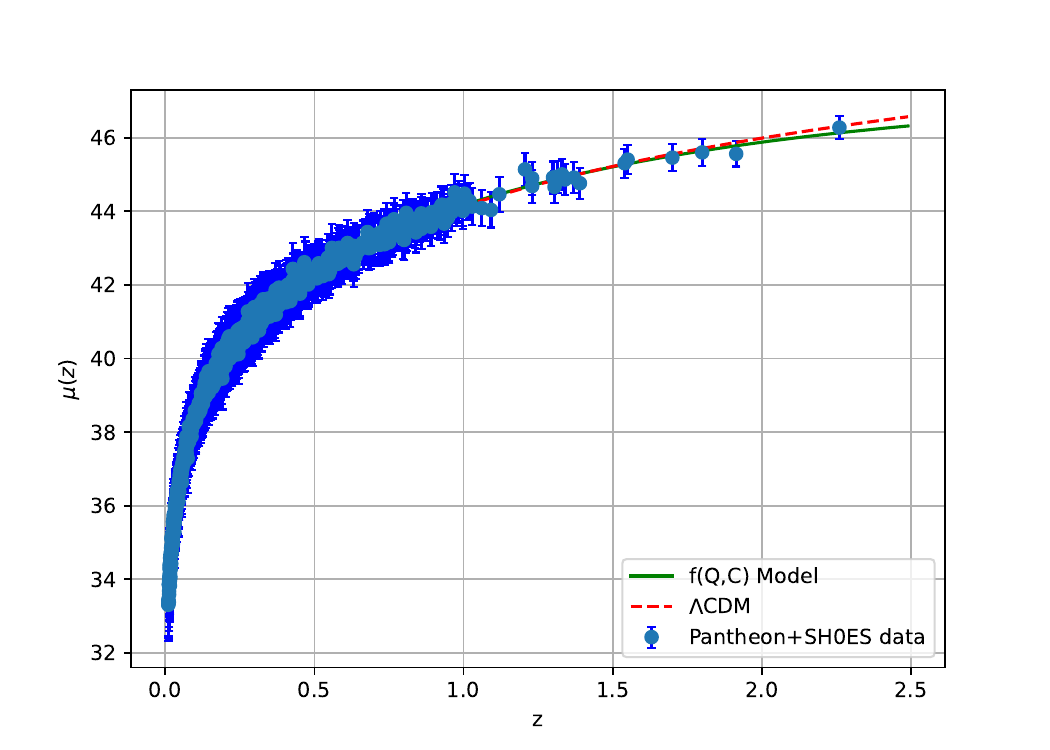} 
    \caption{Error bar plots for 1701 data points from the $Pantheon + SHE0ES $ datasets together with best-fit plots.}
    \label{III} 
\end{figure}

\begin{figure}[ht!]
    \centering
    \includegraphics[width=0.9\textwidth]{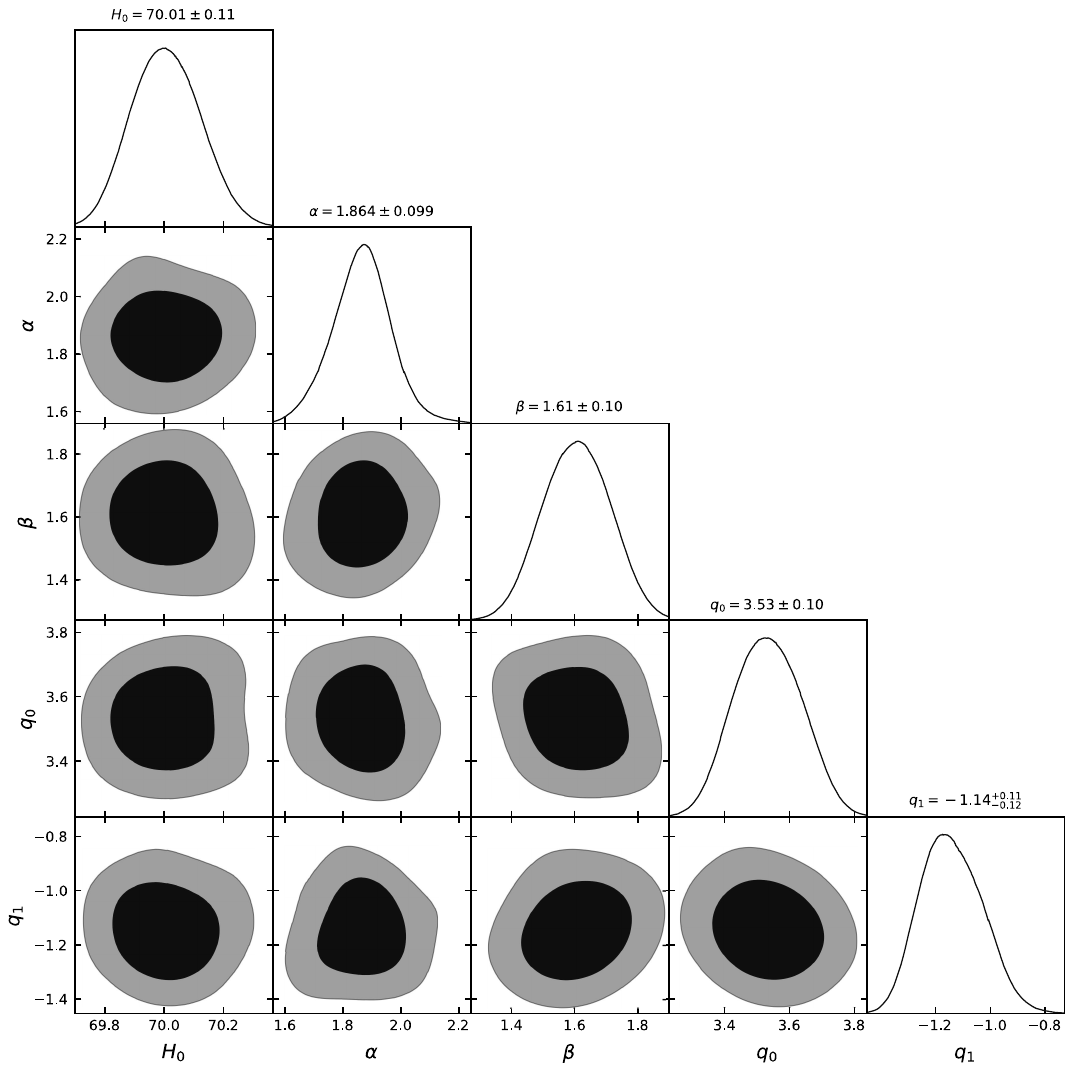} 
    \caption{The MCMC confidence contours derived from constraining the \( f(Q,C) \) model via $Pantheon+SH0ES$ dataset are shown in the plot.}
    \label{IV} 
\end{figure}

The MCMC analysis was performed using the same number of steps and walkers as in the CC example. Figure \ref{III} presents the distance modulus profile, whereas Figure \ref{IV} displays the \( 1\text{-}\sigma \) and \( 2\text{-}\sigma \) confidence level contour plots. The model shows excellent agreement with the $Pantheon+SH0ES$ dataset. The marginal values of all model parameters obtained by using  this dataset are listed in Table \ref{tab:I}.

\begin{table}[h!]
    \centering
    \begin{tabular}{c c c}
    \hline
       Parameters  & $OHD$ & $Pantheon+SH0ES$  \\
       \hline
         $H_0$ & $70.01 ^{+0.12} _{-0.11}$ & $70.01 \pm 0.11$ \\
         $\alpha$ & $1.863 \pm 0.087$& $1.864\pm 0.099$\\
         $\beta$ & $1.58 \pm 0.11$ & $1.61 \pm 0.10$ \\
         $q_0$ & $3.53\pm 0.11$ & $3.53\pm0.10$ \\
         $q_1$ & $-1.151 \pm 0.094$ & $-1.14^{+0.11}_{-0.12}$\\
         \hline
    \end{tabular}
    \caption{Constrained values of the model parameters obtained from the $OHD$ and $Pantheon+SH0ES$ data samples.}
    \label{tab:I}
\end{table}

\begin{table}[h!]
    \centering
    \begin{tabular}{c c c c}
    \hline
      Datasets   & $q_0$ & $z_t$ & $\omega_0$  \\
    \hline
      $OHD$ & $-0.2815^{+0.25}_{-0.15}$ & $0.98^{+0.007}_{-0.066}$& $-0.545^{+0.49}_{-0.66}$ \\
      $Pantheon+SH0ES$ & $-0.2584^{+0.26}_{-0.28}$ &$0.76^{+0.012}_{-0.048}$ &$-0.695^{+0.46}_{-0.56}$ \\
      \hline
         & 
    \end{tabular}
    \caption{Best-fit values of the cosmological parameters and statistical analysis results for the $OHD$ and $Pantheon+SH0ES$ datasets, including confidence levels.}
    \label{tab:II}
\end{table}

\newpage
\section{Cosmographic Parameters}
\label{section_5}
\subsection{Deceleration Parameter}

The deceleration parameter measures the rate at which the expansion of the universe changes over $t$, or $z$. A positive $q(z)$ indicates a decelerated expansion, typically associated with matter or radiation dominance. In contrast, a negative value indicates accelerated expansion, as seen in the present DE-dominated universe. Tracking its behavior helps to understand the impact of different cosmic components on the expansion of the universe.  
The equation for \( q(z) \), obtained from the parametrically derived \( H(z) \) (Eqn. \eqref{30}), is given by:
\begin{equation}\label{36}
    q(z)=\frac{1}{(1+z)(\alpha-1)}\Big[(1+z)\big[2+q_0(\alpha-1)^2+\alpha(\alpha+q_1-3)-q_1(\alpha-1)^2\beta\big]+(q_1-\alpha q_1)\log[\alpha+z]\Big].
\end{equation}
\begin{figure}[H]
    \centering
    \includegraphics[width=0.5\textwidth]{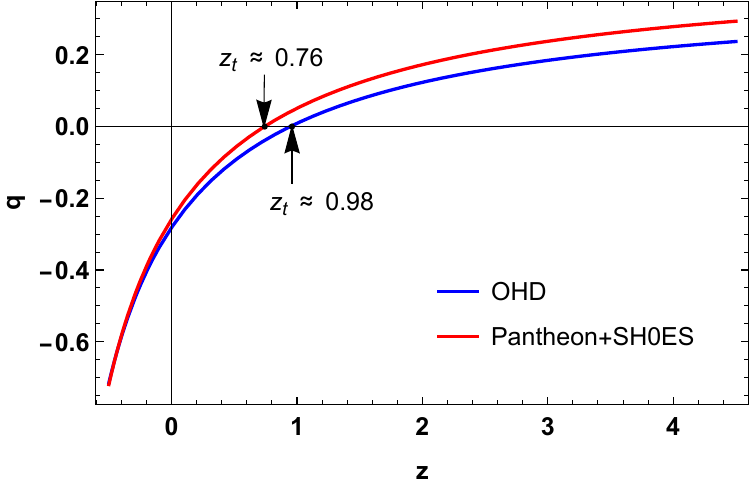} 
    \caption{Plot of deceleration parameter $q(z)$ versus redshift $z$.}
    \label{V} 
\end{figure}
Figure \ref{V} depicts the evolution of the universe, highlighting its transition from early-time deceleration to late-time acceleration, as determined by the constrained values of the model parameters derived from the observational data used in this article. The transition redshifts are obtained as $z_t\approx 0.98$ and, $z_t\approx0.76$ for $OHD$ and $Pantheon+SH0ES$ datasets, respectively. The $OHD$ indicates a delayed transition for cosmic acceleration. The current values of the deceleration parameter for the  $OHD$ and $Pantheon+SH0ES$ samples are observed to be $q_0\approx-0.30$ and $q_0 \approx -0.25$, respectively. This confirms that the universe is undergoing an accelerated expansion, with the negative values highlighting the dominance of DE in the current epoch. From these values, it is evident that the model initially struggles to align with the widely accepted value of \( q_0 \approx -0.55 \). However, as it evolves, the model may successfully transition into an accelerating de Sitter regime.

It is necessary to note that a fundamental feature of any viable cosmological model is its ability to describe the transition from an early decelerating phase, which allows for structure formation, to the current accelerated expansion, as indicated by observational data. In $f(R)$ gravity, this transition arises naturally through modifications to the curvature scalar, which induce an effective dark energy component \cite{mandal2025dynamical}. Similarly, $f(T)$ gravity achieves this transition by modifying the torsional properties of spacetime, leading to a self-accelerating behavior without the need for an explicit cosmological constant \cite{mishra5100074exploring} but in the present work, the transition in $f(Q, C)$ gravity is driven by the interplay between the nonmetricity scalar $Q$ and the boundary term $C$, which together govern the dynamical evolution of the expansion rate. This approach ensures a smooth and observationally consistent transition between different cosmic phases. The key distinction of $f(Q, C)$ gravity lies in its geometric foundation, where cosmic acceleration emerges from nonmetricity rather than curvature or torsion. The results obtained here further support the idea that extended gravity models can successfully account for the observed cosmic dynamics while offering new insights into the underlying nature of gravitational interactions.

\subsection{Density and Pressure}
With the help of Eqns. \eqref{23}, \eqref{24} and \eqref{31}, the energy density and pressure for the derived DE model are given as

\begin{equation}\label{37}
    \rho_{DE}=\frac{1}{\kappa}\Big[3H^2 + \gamma_1 (n-0.5)(-6H^2)^n\Big],
\end{equation}

\begin{equation}\label{38}
    p_{DE}=-\frac{1}{\kappa}\Big[2\dot{H}+3H^2 + \gamma_1 (n-0.5)(-6H^2)^n + \frac{\gamma_1 n(2n-1) \dot{H}}{3} (-6)^n(H^2)^{n-1}     \Big].
\end{equation}

The value of $\gamma_1$ determines how $\rho_{DE}$ and $p_{DE}$ are expressed. This suggests that the nonmetricity factor $Q$ directly impacts the obtained model. To preserve a positive energy density and the accelerating features of the EoS parameter, we then set the values of our model parameters $\gamma_1$ and $\gamma_2$, appropriately. Furthermore, an integer value of $n>1$ is required to achieve valid results for noninteger values, as the model does not correlate correctly. The model parameters, $q_0$, $q_1$, $\alpha$, and $\beta$, affect the energy density and pressure of the DE model. Therefore, we use $\gamma_1=0.235$ and  $n=2$, to keep the Hubble and deceleration parameters within the ranges suggested by cosmological discoveries.

\begin{figure}[H]
    \centering
    \includegraphics[width=0.495\textwidth]{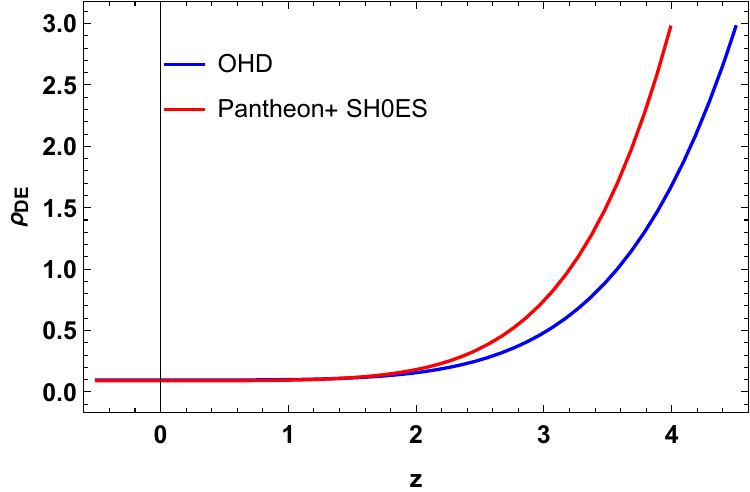} 
    \caption{Plot of energy density $\rho$ versus redshift $z$.}
    \label{VI} 
\end{figure}

Figure \ref{VI} presents the evolution of the DE density \( \rho_{DE} \) as a function of redshift \( z \). At low redshifts (\( z \approx 0 \)), the results from both datasets demonstrate similar trends in energy density, which approach near zero. This behavior aligns with the current epoch of accelerated cosmic expansion, which is dominated by DE or low-density matter. At higher redshifts, the energy density \( \rho_{DE} \) exhibits exponential growth, with the $Pantheon+SH0ES$ dataset predicting slightly higher values than the $OHD$ dataset. This indicates a denser universe in the past, particularly for redshifts (\( z > 2 \)). As the redshift approaches (\( z \to -1 \)), DE density \( \rho_{DE} \) appears to stabilize or entirely dominate, in agreement with the theoretical expectations of an accelerating de Sitter universe. 

\begin{figure}[h!]
    \centering
    \includegraphics[width=0.5\textwidth]{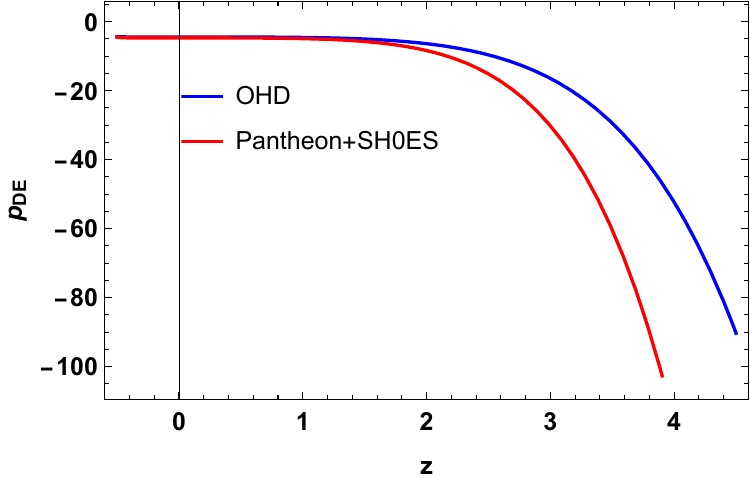} 
    \caption{Plot of pressure $p$ versus redshift $z$.}
    \label{VII} 
\end{figure}

Figure \ref{VII} shows the variation in DE pressure \( p_{DE} \) as a function of redshift \( z \) for two datasets: $OHD$ and $Pantheon+SH0ES$. The plot illustrates the evolution of pressure across different cosmic epochs. In the past (\( z > 0 \)), corresponding to the early universe, the pressure was highly negative, with the $Pantheon+SH0ES$ dataset showing a steeper decline than the $OHD$ dataset. This indicates a stronger influence of DE in the $Pantheon+SH0ES$ dataset during earlier times, suggesting a more rapid expansion of the universe. At present (\( z \approx 0 \)), both datasets converge to less negative pressure values, which is consistent with the current accelerated expansion dominated by DE and low-density matter. Looking toward the future (\( z \to -1 \)), although the figure does not explicitly extend into this regime, the pressure is expected to stabilize near zero or remain negative, aligning with the universe transitioning into a de Sitter phase. 

\subsection{The EoS Parameter}

\begin{figure}[H]
    \centering
    \includegraphics[width=0.5\textwidth]{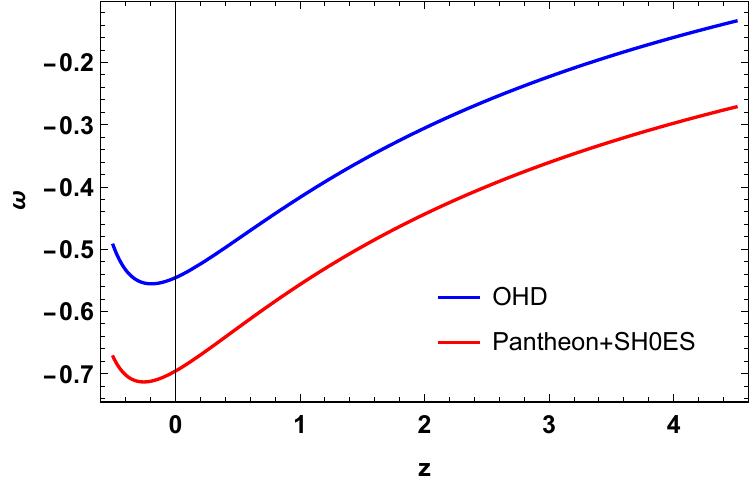} 
    \caption{Plot of the EoS parameter $\omega$ versus redshift $z$.}
    \label{VIII} 
\end{figure}

The EoS parameter $\omega$, which is defined as $\omega=\frac{p_{DE}}{\rho_{DE}}$, is useful to classify the acceleration and deceleration behavior of the universe. To accelerate the universe, the EoS classifies three possible states: the cosmological constant $(\omega = -1)$, phantom $(\omega <- 1)$ era, and the quintessence $(-1 < \omega < -\frac{1}{3})$ era. Figure \ref{VIII} explains the trajectory of the EoS parameter. It indicates that $\omega$  ranges from $-1$ to $- \frac{1}{3}$, throughout the evolution, which means that the whole trajectory lies in the quintessence era. The current values of the EoS parameter for $OHD$ and $Pantheon+SH0ES$ are observed to be $\omega_0\approx -0.55$ and $\omega_0\approx -0.70$, respectively.

\subsection{$r-s $ Parameter}

It is well known that DE promotes the expansion of the universe. Understanding the origins and basic characteristics of DE has gained attention in recent decades.
Consequently, many DE models have surfaced, underscoring the need to make quantitative and qualitative distinctions between them. To distinguish between several DE theories, \textit{Sahni et al.} \cite{sahni2003statefinder} presented a statefinder diagnostic technique. The statefinder parameter $\{r, s\}$, a pair of geometrical parameters used in this approach, is specified as follows:

\begin{equation}
    r=2q^2+q+(1+z)\frac{dq}{dz},
\end{equation}
\begin{equation}
    s=\frac{r-1}{3(q-\frac{1}{2})}.
\end{equation}

\begin{figure}[h!]
    \centering
    \includegraphics[width=0.5\textwidth]{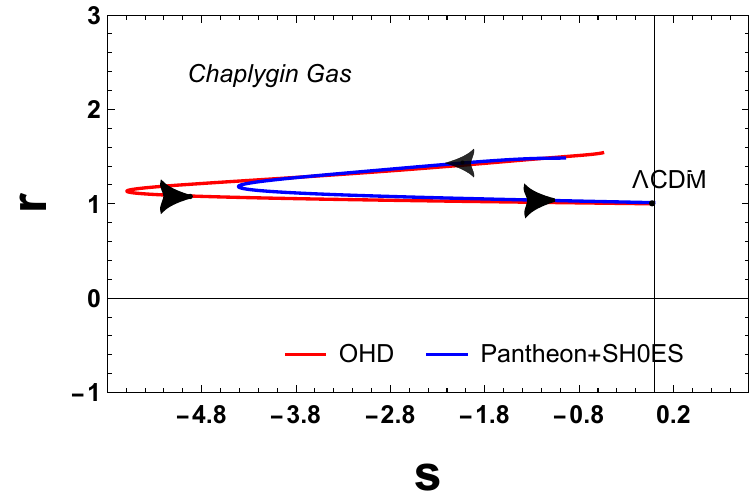} 
    \caption{Plot of $r-s$ parameter versus redshift $z$.}
    \label{IX} 
\end{figure}

The DE models are represented by different values of $(r, s)$; for example, the $\Lambda CDM$ model is represented by $(r = 1, s = 0)$, the Chaplygin gas region is represented by $(r > 1, s < 0)$, and the Quintessence region is represented by $(r < 1, s > 0)$. Figure \ref{IX} illustrates the trajectory of the $r-s$ parameter for $OHD$ and $Pantheon+SH0ES$ datasets. This indicates that $r-s$ the pair falls within the Chaplygin gas regime $(r > 1, s < 0)$ and eventually converges to the $\Lambda CDM$ point $(0, 1)$.

\subsection{$Om(z)$ Diagnostics}
\begin{figure}[h!]
    \centering
    \includegraphics[width=0.5\textwidth]{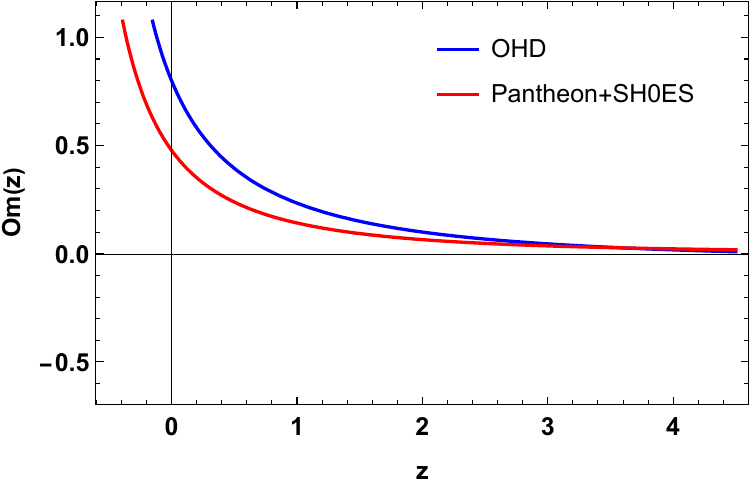} 
    \caption{Plot of Om diagnostic $Om(z)$ versus redshift $z$.}
    \label{X} 
\end{figure}

The Hubble parameter $H$ and the redshift $z$ provide the geometrical diagnosis known as $Om$. It can distinguish between a dynamic DE model and $\Lambda CDM$, both with and without matter density. The negative slope of $Om(z)$ indicates that DE behaves like quintessence $(\omega > -1)$, and the positive slope suggests that DE is a phantom $(\omega < -1)$. Following {\textit{Zunckel}  \& \textit{Clarkson} \cite{zunckel2008consistency} and {\textit{Sahni et al.} \cite{sahni2008two}, $Om(z)$ for a spatially flat universe is defined as
\begin{equation}
    Om(z)=\frac{[H(z)/H_0]^2-1}{(1+z)^3-1}.
\end{equation}

Figure \ref{X} shows that $Om(z)$ has a negative slope, which denotes the quintessence-like behavior of DE in a slowly evolving EoS for both datasets.

\section{Discussion}
\label{section_6}
It is important to mention that model-independent techniques reconstruct cosmic evolution using observational data without assuming an explicit gravitational model. The approach discussed in \cite{capozziello2022model} utilizes such a reconstruction to recover a cosmological model similar to those derived in nonmetric gravity theories, including $f(Q, C)$ gravity. Our analysis in $f(Q, C)$ gravity follows a different approach by assuming a nonlinear parameterization of the deceleration parameter, allowing for a direct connection between theoretical predictions and observational constraints. Unlike model-independent methods, which rely purely on empirical data fitting, the $f(Q, C)$ framework provides a theoretical motivation by modifying the gravitational interaction through the nonmetricity scalar $Q$ and the boundary term $C$. A key comparison is in the evolution of the deceleration parameter. In model-independent reconstructions, $q(z)$ is derived from trends in observational data, often exhibiting flexibility but lacking direct constraints from modifications in fundamental gravity. In contrast, our model was parameterized as Eqn. \eqref{27}, yields a smooth transition from a decelerated phase to an accelerated phase, with transition redshifts $z_t \approx 0.98$ $(OHD)$ and $z_t \approx 0.76$ $(Pantheon+SH0ES)$, consistent with observational findings. This demonstrates that the parameterization within $f(Q, C)$ gravity effectively captures late-time cosmic acceleration while maintaining a connection with fundamental gravitational modifications.

Additionally, model-independent approaches face challenges in predicting the evolution of DE components and distinguishing between different gravity theories. The presence of the boundary term $C$ in our framework introduces additional dynamical effects that are absent in purely empirical reconstructions. This highlights the advantage of our method in providing a more structured theoretical foundation for cosmic evolution. In short, while model-independent techniques are valuable for empirical validation, our approach in $f(Q,C)$ gravity offers a complementary perspective by embedding observationally consistent behavior within a modified gravity framework. Future studies may benefit from combining model-independent reconstructions with specific gravity models to enhance our understanding of cosmic acceleration.

\section{Conclusions}
\label{section_7}
The present theory of accelerating expansion of the universe has become more exciting with time. To find a good representation of the accelerating universe, many dynamic DE models and modified gravity theories have been used in different ways. In this work, we employ an extension  $f(Q)$ 
gravity, along with a boundary term $C$, that is, $f(Q,C)$ gravity. The nonlinear functional form of $f(Q,C)$ gravity is given by Eqn. \eqref{31}. We execute the parameterization of the logarithmic deceleration parameter (Eqn. \eqref{27})  with the help of $f(Q,C)$ gravity in the FLRW universe. The optimal results are determined with the help of  \( \chi^2 \) minimization method to identify the best-fit values for the model parameters \ $\alpha$, $\beta$, $q_0$, and $q_1$. This process involves the use of data samples, mainly $31$ data points from the CC dataset for Hubble measurement and $1701$ data points from the $Pantheon+SH0ES$ dataset for SNeIa. Table \ref{tab:I} presents the values of the constraint parameters along with their corresponding 1-\( \sigma \) confidence intervals. Additionally, Table \ref{tab:II} provides the best-fit values of the cosmological parameters for the current epoch. We compute and study cosmographic parameters, such as the deceleration parameter, pressure, energy density of the DE model, EoS parameter, statefinder parameter, and $Om$ diagnostic.

The results of this study highlight the effectiveness of $f(Q, C)$ gravity in describing the expansion history of the universe without relying on scalar fields or exotic dark energy components. Like other modified gravity theories, such as $f(R)$ and $f(T)$ gravity, the framework explored here provides a purely geometric mechanism to achieve late-time acceleration \cite{das2025perfect}.
Figure \ref{I} shows the trajectory of the Hubble parameter, which indicates that it aligns well with the standard $\Lambda CDM$ model. The current values of the Hubble parameter $H_0$ for $OHD$ and $Pantheon+SH0ES$ data samples are $H_0= 70.01 ^{+0.12} _{-0.11} kms^{-1}Mpc^{-1}$ and, $H_0=70.01 \pm 0.11 kms^{-1}Mpc^{-1}$ respectively, which are similar to the results of \cite{bhoyar2024resolving,gadbail2022parametrization}. The behavior of the deceleration parameter $q(z)$ is shown in Fig. \ref{V}. At higher redshifts, where \( q(z) \) has positive values, the model exhibits a deceleration phase. Upon crossing the transition redshift values, \( z_t = 0.98^{+0.007}_{-0.066} \) for the $OHD$ dataset and \( z_t = 0.76^{+0.012}_{-0.048} \) for the $Pantheon+SH0ES$ dataset, the model transitions into an accelerated phase at lower redshifts. As the model continues to evolve, it is expected to successfully transition into an accelerating de Sitter phase soon. The transition from deceleration to acceleration observed in this model aligns well with the predictions of alternative theories, reaffirming that modifications to the gravitational sector can account for cosmic acceleration \cite{weber2025inflation}. The current values of the deceleration parameter for each dataset are provided in Table \ref{tab:II}. These results align closely with the arguments presented in \cite{mamon2017parametric,samaddar2024constraining}.

The energy density \(\rho_{DE}\) experiences a steady decline from its high values in the early universe (\( z > 0 \)) to nearly zero as we approach the future (\( z < 0 \)) (Fig. \ref{VI}). This behavior corresponds with the transition from a radiation-dominated era to a vacuum-dominated, de Sitter-like phase, highlighting the model's effectiveness in accounting for the diminishing influence of matter and radiation as the universe expands. These findings are consistent with the reasoning described in \cite{gadbail2022parametrization,solanki2022accelerating,bhardwaj2022evaluation}. 

In parallel, the DE pressure \(p_{DE}\) shifts from significantly negative values in the early universe to values near zero in the later stages, as depicted in Fig. \ref{VII}. This trend mirrors the dynamics of the accelerated expansion promoted by DE. The increasing negativity of pressure further underscores the importance of \( f(Q, C) \) gravity in representing the repulsive forces essential for cosmic acceleration. These results support similar conclusions as those reported in \cite{singh2021f,dixit2020rhde}.

The Equation of the State parameter (Fig. \ref{VIII}) exhibits a negative behavior, indicating that it lies within the quintessence era. This behavior suggests that the present universe is undergoing an accelerating phase, reinforcing the notion that DE plays a significant role in its dynamics. It does not cross the phantom divide line for $(z<0)$ for each dataset. The current values of the Eos parameter are $\omega_0=-0.545^{+0.49}_{-0.66}$ for $OHD$ and $\omega_0=-0.695^{+0.46}_{-0.56}$ for $Pantheon+SH0ES$ data samples. These results align closely with the arguments presented in \cite{ghaffari2022kaniadakis,bhoyar2024generalized,bhoyar2024cosmic}.

The statefinder \( r-s \) parameter for constrained values of the model parameters, derived from $OHD$ and $Pantheon+SH0ES$ data samples, is presented in Fig. \ref{IX}. Initially, the trajectory of the \( r-s \) plane is located in the region where \( r > 1 \) and\( s < 0 \), which is typically associated with the Chaplygin gas region. As the model evolves, the pair \( (r,s) \) converges to \( (1,0) \), aligning it with the widely accepted \( \Lambda CDM \) model. These results agree with the results highlighted in \cite{gadbail2022generalized,al2021statefinder,gorini2003can}.

In Figure \ref{X}, the evolution of the $Om(z)$ diagnostic is depicted with a negative slope. This suggests that our model aligns with the quintessence phase of DE for each dataset, characterized by a slowly evolving equation of state. This finding is consistent with theoretical expectations for quintessence, where DE is dynamic and evolves in contrast to the static nature of the cosmological constant. These results correspond well with the arguments discussed in \cite{gadbail2021viscous,bhoyar2024generalized,pradhan2021barrow}.

To summarize, the choice of \( q(z) \) (Eqn. \ref{27}) with a logarithmic term is somewhat arbitrary and is adopted here to explore the impact of the logarithmic term on the resulting cosmological model and its parameters. This assumption also helps close the system of equations. Since the true nature of the universe remains elusive, parameterizing \( q(z) \) offers a simple yet effective approach to studying the universe's transition from a decelerating to an accelerating expansion phase while paving the way for future investigations into the properties of DE. Incorporating additional observational datasets into this analysis would undoubtedly improve the accuracy of constraints expansion history of the universe, positioning this work as a foundational step in that direction.
Furthermore, the \( f(Q, C) \) gravity framework provides a unified explanation of the physical behavior of the universe in its early, current, and late stages by analyzing the dynamics of isotropic pressure, energy density, stability parameters, and energy conditions. This model offers a comprehensive alternative to \( \Lambda \mathrm{CDM} \). However, the $f(Q, C)$ approach offers a different perspective by using nonmetricity as the driving force, separating it from curvature-based and torsion-based modifications. These findings suggest that extended gravity models provide a compelling avenue for further exploration, potentially leading to a deeper understanding of gravitational physics beyond GR. Future studies could refine this framework by considering additional observational constraints and exploring higher-order extensions of $f(Q, C)$ gravity.

\section*{Data Availability Statement}
There is no new data associated with this article.

\bibliographystyle{rsc}
\bibliography{ref}


\end{document}